\documentclass[doublecol]{epl2} 
\usepackage{graphicx}
\usepackage{epstopdf}
\usepackage{amssymb}
\usepackage{amsmath}
\usepackage{amsthm}
\usepackage[extension=xxx]{hyperref}
\usepackage{units}\usepackage{color}

\def\be{\begin{equation}}   \def\ee{\end{equation}}
\def\eq#1{{eq.~(\ref{#1})}}    \def\fig#1{{fig.~\ref{#1}}}

\def\ron{{r_{\rm on}}}  \def\koff{{k_{\rm off}}}   \def\kon{{k_{\rm on}}}   \def\koffo{{k_{\rm off}^0}}

\title{Rigidity sensing by stochastic sliding friction}
\author{Pierre Sens\inst{1}}
\shortauthor{P. Sens}

\institute{                    
  \inst{1} Laboratoire Gulliver, UMR 7083 CNRS-ESPCI, 10 rue Vauquelin, 75231 Paris Cedex 05 - France}
\date{\today}

\abstract{
The sliding friction force exerted by stochastic linkers interacting with a moving filament is calculated. The elastic properties of the substrate on which the linkers are anchored are shown to strongly influence the friction force. In some cases, the force is maximal for a finite substrate rigidity. Collective effects give rise to a dynamical instability resulting in a stick-slip behaviour, which is substrate-sensitive. The relevance of these results for the motility of crawling cells powered by an actin retrograde flow is discussed.}

\begin{document}

\maketitle

\section{Introduction}
Large-scale biological adhesion often involves a collection of ligand and receptor molecules undergoing stochastic binding and unbinding. The transient nature of cellular adhesion is crucial to such processes as motility, during which the cell must both exert a traction force and slide over a substrate. 
The cytoskeleton interacts with the extra-cellular matrix (ECM) through  transmembrane receptors such as integrin binding to components of the ECM such as fibronectin \cite{bershadsky:2006}.
Crawling cells often form broad and flat protrusions, called lamellipodia, where  polymerisation of actin filaments against the cell membrane and contraction of the actin network by myosin motors result in an actin retrograde flow moving away from the cell leading edge \cite{pollard:2003}. Transient cytoskeleton adhesion to the ECM amounts to an effective friction on the actin retrograde flow that pushes the cell edge forward. Cells can sense various external cues, including the rigidity of their environment \cite{pelham:1997,lo:2000,schwarz:2007}. Rigidity sensing might be in part permitted by the stochastic nature of the friction force. Furthermore, the actin retrograde flow is widely reported to be irregular \cite{jurado:2005}, sometimes displaying periodic oscillations \cite{giannone:2004}. This behaviour may be due to collective effect among the proteins linking the cytoskeleton to the extracellular medium.

The pioneering work of Schallamach \cite{schallamach:1963} showed that the friction force on an object sliding over a substrate covered with microscopic stochastic linkers   can show a non-monotonic behaviour with the sliding velocity. The friction force may decrease with increasing velocity in some range of parameters, a behaviour that is usually associated with dynamical instabilities and stick-slip \cite{rice:1983,persson:1998}. Combination of experimental and analytical works showed that stick-slip  occurs between surfaces coated by surfactant layers \cite{drummond:2003} and between an actin filament and a substrate coated with myosin motors \cite{placais:2009}. Numerical simulations have confirmed the relationships between macroscopic frictional phenomena (including stick-slip) and the dynamics of formation and rupture of microscopic bonds \cite{filipov:2004}. More  recently, this behaviour has been put in the context of cell motility by Chan and Odde \cite{chan:2008,bangasser:2013} and has  been theoretically investigated in depth by several groups \cite{srinivasan:2009,li:2010,sabass:2010}. 

Several models exist for cellular rigidity sensing \cite{schwarz:2013, lelidis:2013}, but the role of substrate elasticity, and in particular the influence of substrate-mediated elastic interactions, on the friction force exerted by stochastic linkers has not yet received a synthetic analytical treatment. Based on computer simulations, Chan and Odde argued that the stochastic traction force could be optimal for intermediate substrate stiffness if the element driving the filament motion (myosin motors in their case) impose a force-dependent filament velocity \cite{chan:2008,bangasser:2013}. We show below that their treatment of the substrate compliance (one large spring connecting all the adhesion molecules together) is insufficient. Furthermore, considering the possible generic role sliding friction might play in rigidity sensing during cell migration, a more general conceptual model is needed. We present here a simple derivation of the force-velocity relationship that qualitatively reproduces the relevant features of the complete solution presented in  \cite{srinivasan:2009,li:2010,sabass:2010}, and has the advantage of being amenable to analytical treatment. 
This expression is then used to analytically derive the stability of  a collection of stochastic linkers interacting with an elastic element and the occurrence of a stick-slip instability. The deformability of the substrate may easily be included in the model. Remarkably, the sliding force for a given sliding velocity is predicted to be maximal for a particular value of the substrate stiffness if the velocity is sufficiently large. 
%The relevance of these results for the traction force powering crawling cells is briefly discussed.

\begin{figure}%  figure placement: here, top, bottom, or page
 \onefigure[width=8.5cm]{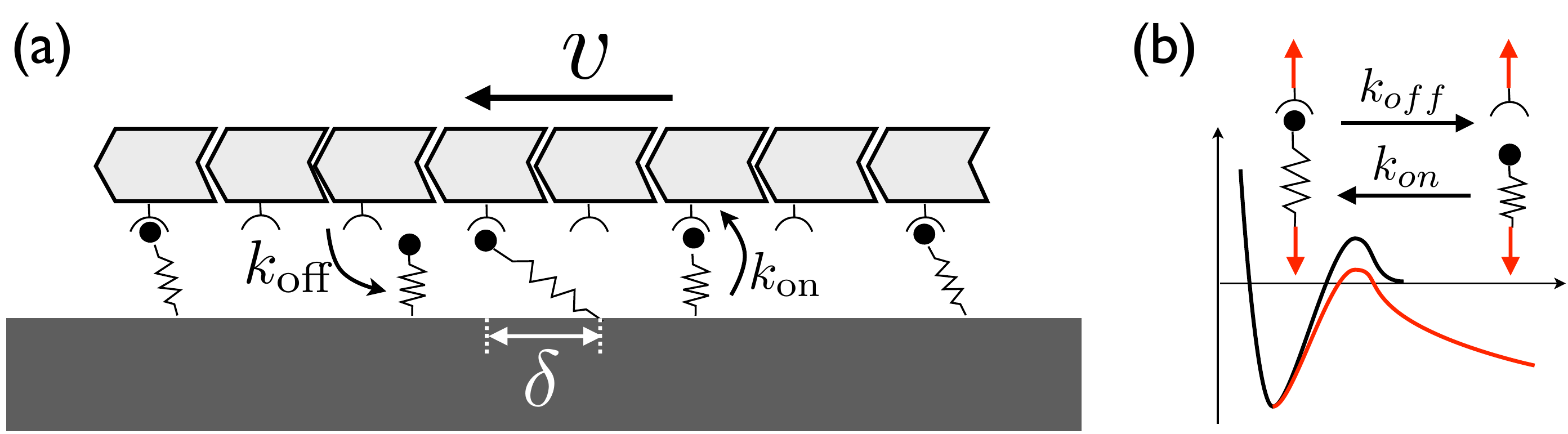}
   \caption{{\em Sketch of the model}: {\bf a)} A filament slides at a velocity $v$ over a substrate covered with hookean molecular linkers with binding and unbinding rates $\kon$ and $\koff$. Bound linkers are stretched by the moving filament, exerting a restoring force linear with their extension,  approximated by the filament's displacement while the linker is bound $\delta$. {\bf b)} Following a Kramer's picture where unbinding corresponds to the passage of an energy barrier, which is lowered in the presence of an external load, the linker's unbinding rate is assumed to depend exponentially of the linker's force.}
 \label{sketch}
\end{figure}

\section{Sliding force for an imposed sliding velocity}

\subsection{Mean-field derivation}

The stochastic friction model is sketched in \fig{sketch}. A filament of length $L$ slides with a velocity $v$ over a substrate coated with a density $\rho$ of elastic linkers
%\footnote{The 1-D density along the filament length}
 of stiffness $k_b$. The linkers bind to and unbind from the filament with rates $\kon$ and $\koff$, respectively.
%The sliding force  at steady state may be approximately obtained as follows; t
The average fraction of linkers attached to the filament is $n=\kon/(\kon+\koff)$. Each linker is bound for an average time $1/\koff$, during which it exerts a force that scales like the linker's extension $\delta$ times its stiffness $k_b$. The linker's extension is assumed to be proportional to the displacement of the moving filament while the linker is bound: $\delta =v/\koff$ (see \fig{sketch}) and the average force per bound linker is approximately: $f=k_b v/\koff$. The average force per filament is $F= Nnf$, where $N=L\rho$ is the number of linkers interacting with the filament:
\be
F=Nk_b v \frac{\kon}{(\kon+\koff)\koff}
\label{force}
\ee
This force is linear with the sliding velocity (constant friction coefficient) if the rates are fixed. We are interested in situations where the rates depend on the force $f$ felt by the linker. Since the filament is moving laterally and remains at the same distance from the substrate, linker's binding may presumably be weakly dependent of the filament velocity. On the other hand, a bound linker is stretched by the  motion of the filament, and we focus on a mechano-sensitive unbinding rate $\koff(f)=\koff^0e^{f/f^*}$ \cite{kramers:1940} (see \fig{sketch}). Here, $\koff^0$ is the unbinding rate under no load, and $f^*$ is a typical force above which the off-rate is strongly affected by the force. 

The relationship between the filament velocity and the off-rate (hence the total force $F$ via \eq{force}) may thus be obtained self-consistently using the force-sensitive off-rate $\koff(f)$ and the off-rate sensitive force  $f=k_b v/\koff$:
%\begin{eqnarray}
%n=\frac{\ron}{\ron+r}\quad;\quad f=f^*\frac{\ron\log{r}}{\ron+r}\cr
%r\equiv\frac{\koff}{\koff^0}\quad;\quad v=v_\beta r\log{r}\quad;\quad v_\beta=\frac{\koff^0f^*}{k_b}
%\label{eqsol}
%\end{eqnarray}
\be
F=Nf^*\frac{\ron\log{r}}{\ron+r}\quad{\rm with}\quad v=v_\beta r\log{r}
\label{eqsol}
\ee
where
\be
r\equiv\frac{\koff}{\koff^0}\quad;\quad \ron\equiv\frac{\kon}{\koff^0} \quad;\quad  v_\beta\equiv\frac{\koff^0f^*}{k_b}
\label{eqsol2}
\ee

The friction force $F$ varies non-monotonously with the velocity (see \fig{fofv}). It displays a maximum force $F_{\rm max}$ for a particular sliding velocity $v^*$, with
\be
F_{\rm max}=Nf^*W\left(\frac{\ron}{e}\right)\ ;\ v^*=v_\beta\ron\left(1+\frac{1}{W\left(\frac{\ron}{e}\right)}\right)
\label{fmax}
\ee
where $W$ is the Lamber $W$ function \footnote{$\lim_{x \to 0}W(x)=x$ and $\lim_{x \to \infty}W(x)=\log(x/\log{x})$} (a.k.a product logarithm function) solution of $x=We^W$. As previously reported \cite{schallamach:1963,drummond:2003,srinivasan:2009,li:2010,sabass:2010}, this remarkable feature is the signature of collective effects among linkers sharing the same load, exerted by the moving filament. 

\subsection{Stochastic theory of linker's dynamics}   
One may go beyond  the qualitative derivation given above and  calculate the stationary sliding force by properly taking into account the probability distribution of binding and unbinding times. Here we restrict ourselves to the description of the stationary sliding force and we neglect finite-size effects. As a consequence, the spatial coordinate has no influence on the linker's state and is absent of the treatment below.

Let's call $p_b(f)df$ the probability density that a bound linker exerts a force $f$ on the filament, and $p_u$ the probability that a linker is unbound. At steady state, these quantities, and  the average force $\langle F\rangle$ exerted by $N$ linkers, satisfy the following equations:
\begin{eqnarray}
\frac{d p_b(f)}{d f}=-\koff(f)p_b(f)\frac{dt}{df}\quad&;\quad p_b(0)=p_u\kon\frac{dt}{df}\cr
\langle F\rangle= N\int_0^\infty df\ fp_b(f)&\label{proba}
\end{eqnarray}
with the normalization: $p_u+\int_0^\infty dfp_b(f)=1$, and with $df/dt=k_b v$. One may easily solve
\eq{proba} for constant rates and show that the average force has the expected form given by \eq{force}.

\begin{figure}%  figure placement: here, top, bottom, or page
 \onefigure[width=8.5cm]{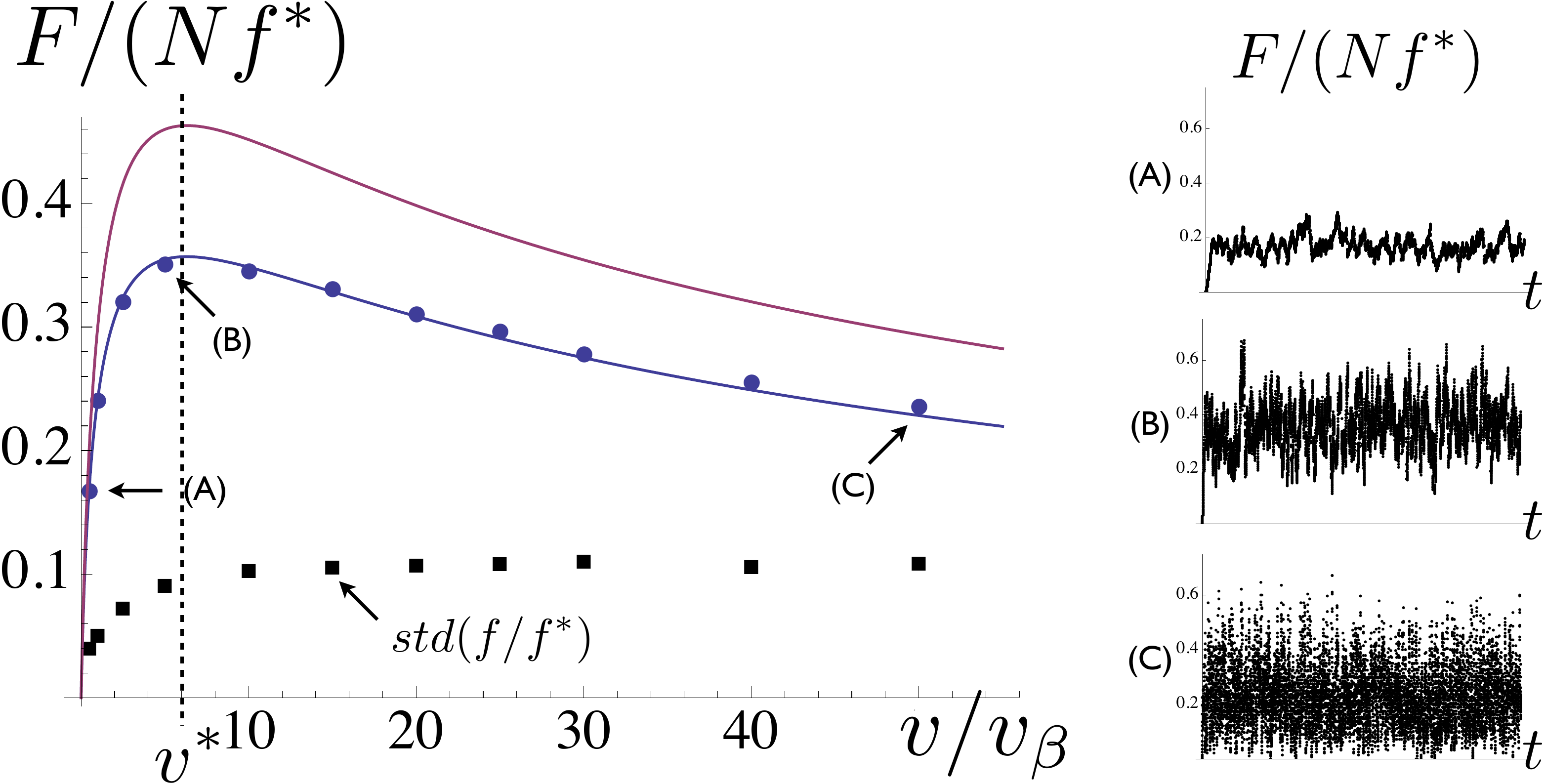}
   \caption{{\em Force-velocity relationship}: Average force caused by a collection of stochastic linkers (with force-sensitive unbinding rates) on a moving filaments, as a function of the filament velocity (with $\ron=2$). Blue dots are the results of simulations \cite{simul-sliding_friction}, the blue curve is the exact solution given by \eq{proba2}, and the red curve is the simplified solution given by \eq{eqsol}. The black squares show the standard deviation of the force. Also shown are three force time-traces obtained from the simulations for low (A), intermediate (B), and high (C) sliding velocities. }
   \label{fofv}
\end{figure}

If the unbinding rate varies exponentially with the linker's force $f$: $\koff=\koffo e^{f/f^*}$, the solution of \eq{proba} reads:
\begin{eqnarray}
p_u=\frac{\tilde v}{\tilde v+\ron e^{1/\tilde v}\Gamma\left[0,\frac{1}{v}\right]}\ ;\ 
p_b(f)=\frac{\ron p_u}{f^*\tilde v}e^{\frac{1}{\tilde v}\left(1-e^{f/f^*}\right)}\cr
\langle F\rangle=Nf^*\frac{\ron e^{1/\tilde v}\int_0^\infty df.f e^{-e^f/\tilde v}}{\tilde v+\ron e^{1/\tilde v}\Gamma\left[0,\frac{1}{\tilde v}\right]}\quad{\rm with}\quad \tilde v\equiv v/v_\beta,
  \label{proba2}
\end{eqnarray}
using the definitions of \eq{eqsol2}, and the incomplete Gamma function   $\Gamma[0,x]=\int_x^\infty dfe^{-f}/f$. This solution \eq{proba2} can be found in earlier publications \cite{srinivasan:2009,li:2010,sabass:2010}.

 As shown in \fig{sketch}, the average force $\langle F\rangle$ given \eq{proba2} agrees perfectly with stochastic simulations. Importantly, it possesses the same qualitative features as the approximate expression given \eq{eqsol}, and in particular the non-monotonicity of the $F(\tilde v)$ relationship. The approximation is valid in the linear regime and correctly locates the sliding velocity $v^*$ at which the friction force reaches a maximum. It overestimates the maximum force by a factor weakly dependent of $\ron$ and varying between $1$ and $1.3$. The discrepancy is due to the fact that the approximation assumes a Poissonian unbinding probability distribution, which is not valid at high force (see \eq{proba2}). 

%In addition to its non-monotonicity, the sliding force, and in particular its 
The fluctuations of the sliding force display interesting statistical features. The standard deviation of the force obtained from stochastic simulations, also shown in \fig{fofv}, continuously increases with the velocity up the velocity  at maximum force $v^*$ and remains constant for $v>v^*$ (but keeps increasing relatively to the decaying force). Variation of the force with time for different velocities show that at high velocity, the system frequently reaches zero force, which corresponds to complete linkers unbinding and free sliding. This is to be expected, as the collection of linkers cannot sustain a force larger than the maximal force $F_{\rm max}$ at steady state. Fluctuations in the number of bound linkers can cause the sliding force to transiently exceed its maximum average value under high sliding velocity, leading to complete linkers unbinding. While the system remains stable on average under the strong constraints of a fixed sliding velocity, it is well known, and shown below, that the flexibility of the pulling device and/or of the substrate can give rise to a dynamical instability in systems displaying a biphasic force-velocity relationship as in \fig{fofv} \cite{persson:1998}.

\section{Sliding friction and rigidity sensing}
The simplified model of \eq{eqsol}, which qualitatively reproduce the relevant dynamical features of the system, permits a complete analytical treatment of the instability.  It is adopted henceforth, and it was checked numerically that all the results given below are also present in the full model leading to \eq{proba2}.
\subsection{ Stick-slip instability for a soft puller}  
\begin{figure}[t] %  figure placement: here, top, bottom, or page
  \onefigure[width=8.5cm]{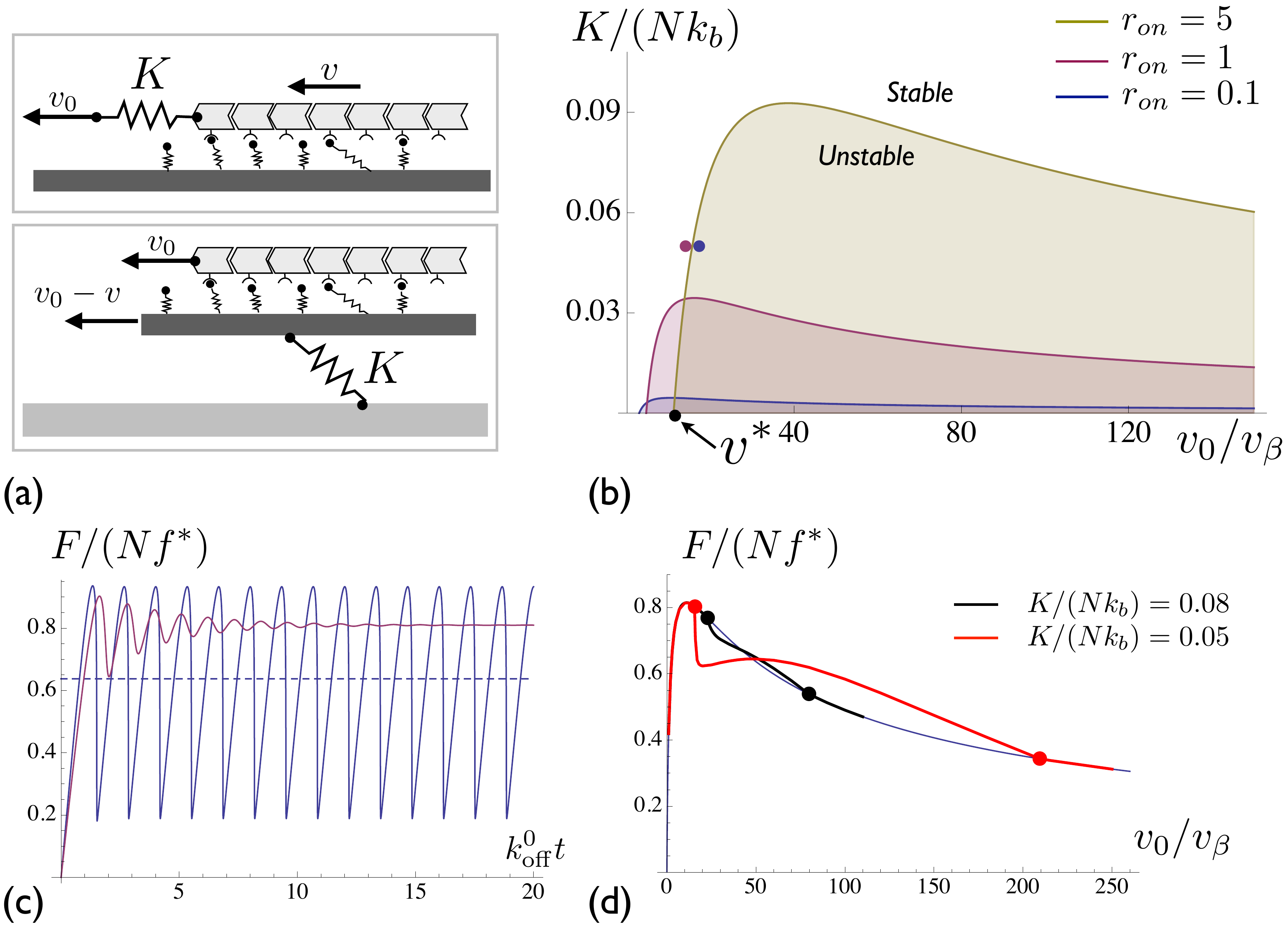} 
   \caption{{\em Filament pulled by a spring}:  {\bf a)} Sketch of a filament attached to an elastic spring of stiffness $K$. One end of the spring is pulled at a constant velocity $v_0$, while the other end is attached to the filament and moves at a velocity $v$. An equivalent dynamical system is obtained if the spring connects the linkers-coated surface to an immobile substrate. {\bf b)} Stability diagram in terms of the normalized pulling velocity $v_0/v_\beta$ and the ratio of spring to linker's stiffness $K/(Nk_b)$, for three values of the normalized on-rate $\ron$. The dots in the phase diagram correspond to the two curves of (c). {\bf c)} Variation of the sliding force with time in the stable (red) and unstable (blue) regimes, for $\ron=5$. In the unstable regime (the shaded regions of (b)), phases of low friction (low bound linkers fraction $n$) alternate with phases of high friction in a  stick-slip fashion. The dashed blue line is the average force in this regime. {\bf d)} Relation between the pulling velocity and the sliding force for two values of the stiffness ratio (for $\ron=5$). The range of velocities corresponding to stick-slip is marked by a pair of large dots, corresponding to the boundaries of the phase diagram (b). The force-velocity curves follows the relationship given in \eq{eqsol} in the stable regime and shows complex nonlinearities in the unstable regime (see text).}
   \label{stickslip}
\end{figure}

The filament velocity is generally not strictly imposed in  experimental situations. For instance, a velocity of a filament pulled by an optical trap moved at constant velocity \cite{placais:2009} may deviate from the imposed velocity due to the softness of the trap. This is mimicked by a spring of controllable stiffness $K$ attached to the filament by one end and pulled at the other end at imposed velocity $v_0$ (\fig{stickslip}). One can easily see that the case of a filament moved at constant velocity over a substrate itself attached elastically to an immobile substrate is stricktly equivalent to the soft puller model (\fig{stickslip}a).
%The compliance of the pulling device ({\em e.g} an optical trap \cite{placais:2009}) can be mimicked by adding a spring of  stiffness $K$ attached to the filament by one end and pulled at the other end at imposed velocity $v_0$ (\fig{stickslip}). One can easily see that the case of a filament moved at constant velocity over a substrate itself attached elastically to an immobile substrate is strictly equivalent to the soft puller model (\fig{stickslip}a).
The soft puller dynamical system reads:
\begin{eqnarray}
\dot F=K(v_0-v)\qquad \dot n=\koff^0(\ron-(\ron+r)n)\cr
{\rm with}\quad F=Nf^*n\frac{v/v_\beta}{r} \quad {\rm and} \quad v=v_\beta r\log r ,
\label{dynsyst}
\end{eqnarray}
which admits \eq{eqsol} with $v=v_0$  as a stationary solution. Linear stability analysis shows that the stationary solution is stable provided the spring stiffness satisfies:
\be
\frac{K}{N k_b}>\frac{r_0(\log{r_0}-1)-\ron}{r_0(\log{r_0}+1)}\frac{\ron}{\ron+r_0}
\label{condk}
\ee
where $r_0$ is defined by $v_0= v_\beta r_0\log{r_0}$. The critical stiffness ratio corresponds to a Hopf bifurcation \cite{guckenheimer:2002} where the real part of the eigenvalues of the Jacobian matrix of the system defined by \eq{dynsyst} vanishes at the stationary point. 

The condition \eq{condk} is shown graphically in \fig{stickslip}b. The limit of a very soft puller ($K\rightarrow0$) is unstable whenever the force-velocity relationship has a negative slope \cite{persson:1998}, that is when $v_0>v^*$. The rigid puller regime ($K/Nk_b\gg1$) is equivalent to imposing the filament velocity, and is stable as discussed previously. There exists a critical ratio $K/Nk_b$, that depends on $\ron$, above which steady sliding is stable for all velocities (\fig{stickslip}).  Below this critical ratio, a dynamical instability is predicted for a finite velocity range. Steady sliding is stable at low {\em and} high velocity. The latter corresponds to linkers being bound for a very small fraction of the time, during which collective effects cannot take place. Numerical solution of \eq{dynsyst} in the unstable regime show that the filament undergoes periodic non-linear oscillations characteristic of a stick-slip mechanism (\fig{stickslip}c). The stick-slip oscillations correspond to a stable limit cycle of the dynamical system of \eq{dynsyst} (supercritical Hopf bifurcation \cite{guckenheimer:2002}). 

In the stable regime, the force-velocity relationship follows the steady-state solution given by \eq{eqsol} and shows a maximum force for $v=v^*$. The average force abruptly drops upon entering the instability regime, as can be understood by looking at the time variation of the force on either sides of the instability boundary (\fig{stickslip}c). The mean of the oscillating force above the threshold (dashed blue line) is clearly lower than the stationary force below the threshold (red line). The amplitude of the force drop decreases upon increasing the stiffness ratio $K/Nk_b$, as shown in \fig{stickslip}d, and vanishes continuously when the unstable regime disappears. 

\subsection{Sensitivity to substrate stiffness} Our primary interest is to compute the extent to which the stochastic sliding friction sketch in \fig{sketch} is sensitive to the rigidity of the substrate. Let's consider a filament sliding at constant velocity $v_0$ over a semi-infinite elastic substrate of Young's modulus $E$ covered with linkers. The force exerted by a linker bound to the moving filament creates a deformation  field in the elastic substrate that extends far from the linker's anchoring point. 
%(the deformation decays as the inverse distance from the anchoring point \cite{landau_elasticity}). 
The substrate deformation underneath a bound linker (labeled $i$) is thus the sum of the deformation  $u_i$  caused by the force exerted by the linker itself and the deformation $\bar u_i$ caused by all the other bound linkers.  Calling $G_{ij}$ the Green's function  giving the surface deformation at a position $r_i$ on the substrate surface  due to a point force $f_j$ at a position  $r_j$ on that surface, we have \cite{landau_elasticity}:
\be
u_i=G_{ii} f_i\ ,\  \bar u_i=\sum_{j\ne i}^NG_{ij} f_j\quad;\quad G_{ij}\propto \frac{1}{E|r_i-r_j|}
\label{green}
\ee
%Here, $f_i$ is the force on the i$^{th}$ linker and $G_{ij}$ is the Green's function giving the surface deformation at a position $r_i$ on the surface of a  semi-infinite elastic medium of Young's modulus $E$, due to a point force at a position  $r_j$ on that surface \cite{landau_elasticity}. 
In this paper, we restrict ourselves to a mean-field treatment. We do not account for spatial correlation between stickers, nor do we incorporate  finite-size effects. In this case, the forces and deformations of all linkers are identical; $\forall j$: $f_j\equiv f$, $u_i\equiv u$ and $\bar u_i\equiv \bar u$. The deformations due to a linear array of $N$ equally spaced point forces of total length $L$ are: 
\be
u\propto \frac{f}{Ea}\quad,\quad \bar u\propto \sum_{j\ne i}^N\frac{Nf}{EL|i-j|}\simeq \frac{Nf}{EL}\log{N}
\label{deform}
\ee
where numerical prefactors have been dropped for simplicity. An elastic cut-off size $a$ must be introduced in $u$ to account for the fact that forces is not really point-like and to prevent divergencies. This molecular size is expected to be of order of the order the linker's anchor size.

The two deformations have quite different dynamics. The deformation $u$ caused by the local force  follows the kinetics of individual linkerÕs attachment and detachment.
It has a lifetime $1/\koff$ and it acts in series with the linkers extension $\delta$. One may thus define the velocity $v=(\delta+u)\koff$ that accounts  for both deformations, to which is associated an effective linker's stiffness $k_b'$ corresponding to the two springs (of stiffness $k_b$ and $Ea$) in series. On the other hand, the deformation $\bar u$ caused by the other linkers is an ensemble average that is insensitive to individual binding/unbinding events. It effectively amounts to the deformation of a spring $K'=Nf/\bar u$ that couples all the linkers together, and is similar to the  case of a soft puller with a rate of extension $\dot{\bar u}=v_0-v$, as in \eq{dynsyst}. 
 Using \eq{deform}, one finds:
\be\qquad
k_b'= \frac{k_bEa}{k_b+Ea}\quad,\quad K'\propto \frac{E L}{\log{(Nn)}}
\label{effective}
\ee

\begin{figure} %  figure placement: here, top, bottom, or page
  \onefigure[width=8.5cm]{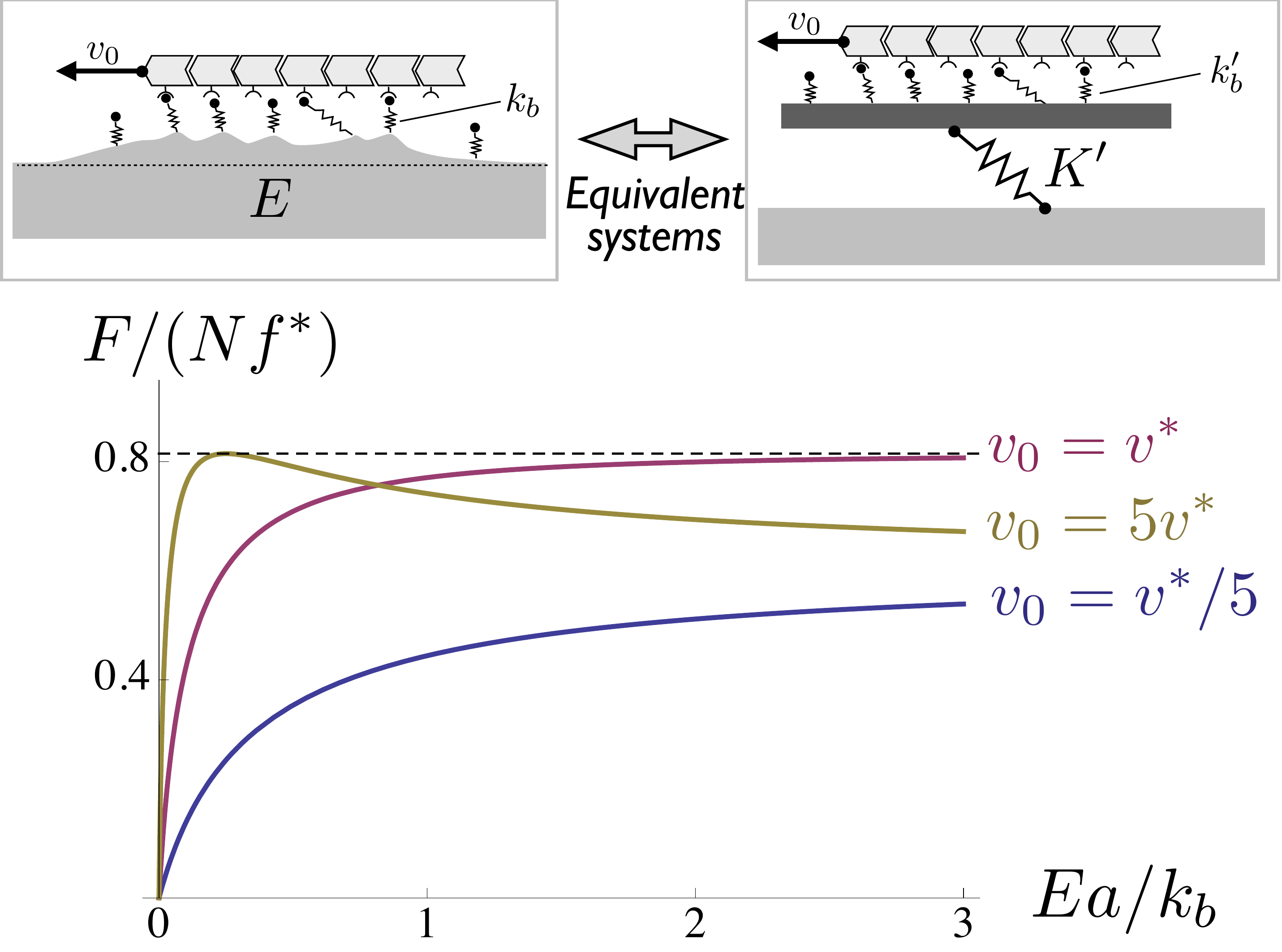} 
   \caption{{\em Variation of the friction force with the substrate's elasticity}: Top: An elastic substrate modifies the effective rigidity of the molecular linker's attaching to the filament ($k_b\rightarrow k_b'$), and introduces an effective spring (of stiffness $K'$) that couples all the linkers together (see text and \eq{effective}). Bottom: variation of the sliding force with the substrate Young's modulus $E$ for three values of the imposed sliding velocity $v_0$ (with $\ron=5$). If $v_0>v^*$ (\eq{fmax}),  the optimal friction force is obtained for a finite substrate elasticity (\eq{estar}).}
   \label{elast}
\end{figure}

The substrate rigidity thus impacts the sliding friction force on two levels. Locally,  it  reduces the apparent stiffness  of the adhesion molecules ($k_b\rightarrow k_b'$), and this modifies the stationary force given by eqs. (\ref{eqsol},\ref{eqsol2}). Globally, it enhances collective effects that may lead to a stick-slip instability. Previous works only discuss the latter effect \cite{chan:2008,bangasser:2013}, but the present analytical model shows that the two effects cannot be dissociated, as the relevant criteria for the instability involves the ratio $K'/(Nk_b')$. The stability criterion \eq{condk} must be modified to account for substrate deformability, because the effective stiffness $K'$ in \eq{effective} depends on time through the fraction $n(t)$ of bound linkers. The stability condition \eq{condk} becomes:
\begin{eqnarray}
\frac{EL}{k'_bN\log N}>\left(1+\frac{\log{n_0}}{\log N}\right)n_0\times\hspace{2cm}\nonumber\\
\hspace{2cm}\frac{r_0\left(\log{r_0}\left(1+\frac{1}{\log{Nn_0}}\right)-1\right)-\ron}{r_0(\log{r_0}+1)}\hspace{1cm}
\label{condk2}
\end{eqnarray}
with $n_0=\frac{\ron}{\ron+r_0}$. This critical value for the stiffness ratio depends on $\ron$ and $r_0$ in a way similar to \eq{condk}, but also (slightly) on $N$.
 
The sliding friction force is  influenced by the substrate stiffness in a non-trivial fashion. Eq. (\ref{fofv}) gives the general force-velocity relation $F(\tilde v)$ in which the  filament sliding speed $v$ only appears in the ratio $\tilde v=v/v_\beta$, where the characteristic velocity $v_\beta$ depends on the linker's rigidity, as given by \eq{eqsol2}.  On non-deformable substrates, a filament with imposed velocity $v_0$ experiences a friction force $F(\tilde v_0)$. On deformable substrates, the effective linker's rigidity is affected by the substrate elasticity ($k_b\rightarrow k_b'$ - \eq{effective}), and the value of the sliding force $F(\tilde v_0')$ is obtained by replacing $\tilde v_0$ by $\tilde v_0'$ in \eq{eqsol}, with  $\tilde v_0'=(k_b'/k_b)\tilde v_0=\tilde v_0\times E a/(k_b+E a)$. Varying the substrate stiffness is thus equivalent to exploring a range of values of $\tilde v$ in  \fig{fofv}, with a corresponding variation of the sliding force.
For a given value of $\tilde v_0$, $\tilde v_0'$ increases linearly with $E$ for very soft substrates and saturates at $\tilde v_0$ for very stiff substrates. For $\tilde v\ll 1$, $F(\tilde v)\sim \tilde v$ so the force is linear with  the substrate stiffness for soft substrate.
Since $F(\tilde v)$, shows a maximum for a particular velocity $\tilde v^*$ of order $\ron$ (\eq{fmax} and \fig{fofv}), the sliding force may also present a maximum for a particular value of the substrate stiffness. 
%For very stiff substrates, the force is $F(\tilde v_0)$, and may be on the increasing of decreasing branch of $F(\tilde v)$ in \fig{fofv}. 
If   $v_0<v^*$, the force increases monotonously with $E$ to a maximum value $F(\tilde v_0)$, obtained on stiff substrates ($E\rightarrow\infty$). If $v_0>v^*$, the force presents a maximum equal to $F_{\rm max}$ (\eq{fmax}) for a particular value of substrate stiffness $E^*$, corresponding to $\tilde v_0'=\tilde v^*$ and given by:
\be
E^*=\frac{k_b}{a}\frac{v^*}{v_0-v^*}\simeq\frac{k_b}{a}\frac{1}{k_bv_0/(f^*\kon)-1}
\label{estar}
\ee
The variation of the friction force with the substrate stiffness is displayed in \fig{elast} for different values of the pulling velocity $v_0$. The friction force is biphasic (first increasing, then decreasing with the stiffness beyond a critical value $E^*$) if $v_0>v^*$.  Using  $\tilde v^*\simeq \ron$ (\eq{eqsol}), the existence of an optimal substrate stiffness for maximal friction is predicted if  $f^*>k_b v_0/\kon$.

\section{Discussion} 
%The results above could be relevant to cell crawling. 
Crawling cells are  able to sense the rigidity of their substrate and are generally thought to perform durotaxis (the tendency to move toward substrates of high stiffness) \cite{disher:2005}. Some cells however exhibit a maximal crawling speed for an optimal substrate stiffness \cite{gardel:2008,stroka:2009}. It is argue in \cite{chan:2008,bangasser:2013} that this can be explained by the stochastic nature of sliding friction, provided the sliding velocity decreases with the sliding force, as could be expected if filament motion is driven by actomyosin contraction. The present model shows that the biphasic variation of the traction force with substrate rigidity is predicted under constant sliding velocity by accurately accounting for the substrate deformation near discrete stochastic linkers.

%Since crawling cells are at least partly powered by the sliding friction of the retrograde actin flow, the present model may provide a physical explanation for this. 

In our model, the  rate of extension of the linker/substrate system is imposed by the filament velocity $v_0$. In the linear regime ($E\ll E^*$ - \eq{estar}), the linkers remain bound to the filament for a fixed amount of time, $1/\koff^0$, regardless of the velocity and the substrate rigidity. The traction force per linker is the product of the linker+substrate extension multiplied by an effective rigidity $k_b'$ that increases with the substrate stiffness (\eq{effective}). Thus in this regime, cells powered by the actin retrograde flow should gain more traction on more rigid substrates. Non-linearities, caused by the increase of the unbinding rate with the tension on the linker, may drastically affect this picture. If the ratio $k_b v_0/(f^*\kon)$ is larger than of order unity, the traction force is maximal on a substrate of rigidity $E^*$ (\eq{estar}), and decreases on stiffer substrates.  The reason for this phenomenon, which could explain the biphasic behaviour observed in \cite{gardel:2008,stroka:2009}, is that although each bound sticker exerts a larger average force on stiff substrates, the fraction of bound stickers decreases with increasing rigidity, leading to a smaller total force. 
Interestingly, it was observed in \cite{gardel:2008} that the typical speed of the actin retrograde flow leading to maximum traction  did not depend on drug treatments affecting actin dynamics or myosin activity. In agreement with this observation, the present model predicts (\eq{fmax}) that while the maximum traction force $F_{\rm max}$ depends on morphological parameters such as the size $N$ of the adhesion zones, the  speed $v^*$ for optimal traction is solely determined by the  linker's molecular parameters and the substrate rigidity.

We investigate whether a variation of substrate rigidity alone can induce a stick-slip transition, as suggested in \cite{chan:2008}. If the filament velocity is strictly imposed ($K'=E/a$), the stability criterion \eq{condk2} shows that the movement should remain stable if $\frac{EL}{k'_b N\log N}=\frac{1}{\rho a}(1+Ea/k_b)$ is larger than a function of $\ron$ necessarily smaller than unity ($a$ and $\rho$ is the linker's anchor size and density along the filament). Since $(\rho a)<1$, we expect the steady motion to be stable for any values of the Young's modulus. This is because one cannot reduce the substrate stiffness without also reducing the linker's effective rigidity. A filament pulled by a soft puller over an elastic substrate may however display Young's modulus-dependent stick-slip. If such a transition occurs, one then expects steady motion over soft substrate and stick-slip over stiffer substrate.

We now estimate whether our results could bear relevance for cytoskeletal flow interacting with  the extracellular matrix. With typical values: 
$f^*=5\unit{pN}$, $\kon=10/\unit{s}$, $\koff^0=1/\unit{s}$ and $k_b=1\unit{pN/nm}$, we find $v_\beta=5\unit{nm/s}$, and a velocity at maximum force $v^*=100\unit{nm/s}$, which is within the velocity range of actin retrograde flow. With a filament of length $L=1\unit{\mu m}$ and a linker density $\rho=1/(20\unit{nm})$, at most $N\sim 50$ linkers may be bound to one filament, and the maximum force that can be reach is $F_{\rm max}=200\unit{pN}$, also within typical cellular-scale forces. If a filament is pulled by an optical trap of stiffness $K=0.1\unit{pN/nm}$, the stiffness ratio is $K/(N k_b)=2\times 10^{-3}$ in \eq{condk}, and one expects to observe stick-slip for pulling velocities $\gtrsim v^*$ (\fig{stickslip}). On deformable substrates, the sliding force should exhibit substrate stiffness sensitivity up to Young's moduli of order $k_b/a=100\unit{kPa}$ (with $a=10\unit{nm}$), which makes it an efficient mechano-sensitive device for most biological tissues. 

Possible extensions of this model could include a mechano-sensitive on-rate and the possibility for a linker to bind at any location along the filament. This can be achieved by introducing a space-dependent bound probability $p_b(x,t)$ in \eq{proba} and by accounting for filament sliding as a convective term in $d p_b/d t$ \cite{srinivasan:2009,li:2010,sabass:2010}. Although this sizably complicates the treatment of the equations, and introduces the linkers rest state as a relevant parameter, it does not qualitatively modify our conclusions.
%, and reduces to the present description is case linkers must stretched to bind to the filament
 Fluid dissipation in the surrounding medium can easily be included in the picture. This adds a linear force $f\propto v$ to the friction force of \fig{fofv}, eventually leading to an increase of force with $v$ for $v\rightarrow\infty$ \cite{drummond:2003,sabass:2010}. Molecular friction should however be much larger than viscous friction, which can only brings small correction to the present picture. It would also be of interest to study the spatial correlations between linkers resulting from substrate-mediated interactions. The mean field model presented here provides an analytical description of rigidity sensing by stochastic friction, but the nature of the dynamical instability might be sensitive to fluctuations and spatial correlations.

%Such instability is shown to be expected for  force and velocity scales relevant to cellular processes. Our work provides a simple an analytical framework within which experimental gliding assays ({\em e.g} based on optical trap, \cite{bormuth:2009,placais:2009}) designed to study the mechano-sensitivity of biological adhesion bonds can be analysed.

\acknowledgments
The work was supported by the Human Frontier Science Program under the grant RGP0058/2011. David Lacoste, Marco Leoni, Ulrich Schwarz and Ken Sekimoto are acknowledged for stimulating discussions.

%\bibliographystyle{unsrt}
%\bibliographystyle{apsrev}
%\bibliographystyle{apsrev4-1}

%\bibliographystyle{eplbib}
%\bibliography{BibTeX_ref_pio.bib}

\end{document}